\documentclass[prd,preprint,tightenlines,superscriptaddress,showpacs,aps,amsmath,amssymb,nofootinbib]{revtex4}
\usepackage{graphicx} 

\newcommand{\beq}{\begin{equation}}
\newcommand{\eeq}{\end{equation}}

\begin{document}

\title{UHE neutrinos from superconducting cosmic strings}

\author{Veniamin Berezinsky}
\email{venya.berezinsky@lngs.infn.it}
\affiliation{INFN, Laboratori Nazionali del Gran Sasso,
                   I--67010 Assergi (AQ), Italy}
\author{Ken D. Olum}
\email{kdo@cosmos.phy.tufts.edu}
\author{Eray Sabancilar}
\email{eray.sabancilar@tufts.edu}
\author{Alexander Vilenkin}
\email{vilenkin@cosmos.phy.tufts.edu}
\affiliation{Institute of Cosmology, Department of Physics and Astronomy,
Tufts University, Medford, MA  02155, USA.}

\begin{abstract}

Superconducting cosmic strings naturally emit highly boosted charge
carriers from cusps. This occurs when a cosmic string or a loop moves
through a magnetic field and develops an electric current. The charge
carriers and the products of their decay, including protons,
photons and neutrinos, are emitted as narrow jets with opening angle
$\theta \sim 1/\gamma_c$, where $\gamma_c$ is the Lorentz factor of
the cusp. The excitation of electric currents in strings occurs mostly
in clusters of galaxies, which are characterized by magnetic fields $B
\sim 10^{-6}$~G and a filling factor $f_B \sim 10^{-3}$. 

Two string parameters determine the emission of the particles: the
symmetry breaking scale $\eta$, which for successful applications
should be of order $10^{9}$--$10^{12}$~GeV, and the dimensionless
parameter $i_c$, which determines the maximum induced current as
$J_{\rm max} =i_c e \eta$ and the energy of emitted charge carriers as
$\epsilon_X \sim i_c \gamma_c \eta$, where $e$ is the electric charge of a
particle. For the parameters $\eta $ and $B$ mentioned above, the
Lorentz factor reaches $\gamma_c \sim 10^{12}$ and the maximum
particle energy can be as high as $\gamma_c\eta \sim 10^{22}$~GeV.
The diffuse fluxes of UHE neutrinos are close to the cascade upper
limit, and can be detected by future neutrino observatories.  The
signatures of this model are: very high energies of neutrinos, in
excess of $10^{20}$~eV, correlation of neutrinos with clusters of
galaxies, simultaneous appearance of several neutrino-produced showers
in the field of view of very large detectors, such as JEM-EUSO, and
$10$~TeV gamma radiation from the Virgo cluster.  The flux of UHE
protons from cusps may account for a large fraction of the observed
events at the highest energies.

\end{abstract}

\maketitle

\section{Introduction}
\label{sec:introduction}

\subsection{Neutrino astronomy}

Ultra-high-energy (UHE) neutrino astronomy at energies above
$10^{17}$~eV is based on new, very efficient methods of neutrino
detection and on exciting theories for neutrino production. The most
interesting range of this astronomy covers tremendously high energies
above $10^{19} - 10^{20}$~eV. In fact, this energy scale gives only
the low-energy threshold, where the new observational methods, such as
space-based observations of fluorescent light and radio and acoustic methods,
start to operate. These methods allow observation of very large areas
and so detection of tiny fluxes of neutrinos. For example the exposure
of the space detector JEM-EUSO \cite{jemeuso} is planned to reach
$\sim 10^6$~km$^2$yr~sr. The upper limits obtained by radio
observations are presented in Fig.~\ref{fig:upper-limits}.

The basic idea of detection by EUSO is similar to the fluorescence
technique for observations of extensive air showers (EAS) from the
surface of the Earth. The UHE neutrino entering the Earth's atmosphere
produces an EAS. A known fraction of its energy, which reaches 90\%,
is radiated in the form of isotropic fluorescent light, which can be
detected by an optical telescope in space.  There is little absorption
of up-going photons, so the fraction of flux detected is known, and
thus EUSO provides a calorimetric measurement of the primary
energy. In the JEM-EUSO project \cite{jemeuso} a telescope with
diameter 2.5 m will observe an area $\sim 10^5$~km$^2$ and will have a
threshold for EAS detection $E_{\rm th}\sim 1\times 10^{19}$~eV.  The
observations are planned to start in 2012--2013.

UHE neutrinos may also be very efficiently detected by observations of
radio emission by neutrino-induced showers in ice or lunar regolith.
This method was originally suggested by G.~Askaryan in the 1960s
\cite{askarian}. Propagating in matter the shower acquires excess
negative electric charge due to scattering of the matter
electrons. The coherent Cerenkov radiation of these electrons produces
a radio pulse. Recently this method has been confirmed by laboratory
measurements \cite{saltz}.  Experiments have searched for such
radiation from neutrino-induced showers in the Greenland and Antarctic
ice and in the lunar regolith. In all cases the radio emission can be
observed only for neutrinos of extremely high energies.  Upper
limits on the flux of these neutrinos have been obtained in the GLUE
experiment \cite{glue} by radiation from the moon, in the FORTE
experiment \cite{forte} by radiation from the Greenland ice, and in the
ANITA \cite{anita} and RICE \cite{rice} experiments from the Antarctic
ice.

Probably the first proposal for detection of UHE neutrinos with
energies higher than $10^{17}$~eV was made in \cite{BS}. It was
proposed there to use the horizontal Extensive Air Showers (EAS) for
neutrino detection. Later this idea was transformed into the
Earth-skimming effect \cite{Fargion} for $\tau$ neutrinos.  Recently
the Auger detector \cite{tau-auger} put an upper limit on UHE
neutrino flux using the Earth-skimming effect (see
Fig.~\ref{fig:upper-limits}).

\subsection{UHE neutrino sources}

What might these new large-area UHE neutrino observatories detect?  On
the one hand, there are without doubt {\em cosmogenic neutrinos},
produced by UHECR particles interacting with the CMB photons.  On the
other hand, there may be neutrinos produced in decays or annihilation
of superheavy particles; this is referred to as the {\em top-down
  scenario}.

Cosmogenic neutrinos were first discussed in \cite{BZ}, soon after the
prediction of the GZK cutoff \cite{GZK}.  There, it was shown that UHE
neutrino fluxes much higher than the observed UHECR flux can be
produced by protons interacting with CMB photons at large redshifts.
The predicted flux depends on the cosmological evolution of the
sources of UHE protons and on the assumed acceleration mechanisms.
Recent calculations of cosmogenic neutrino fluxes (see
e.g. \cite{Kal}--\cite{Sato}) are normalized to the observed UHECR
flux, with different assumptions about the sources.

The energies of cosmogenic neutrinos are limited by the maximum energy
of acceleration, $E_{\rm acc}^{\rm max}$.  To provide neutrinos with
energies above $1\times 10^{20}$~eV, the energies of accelerated
protons must exceed $2\times 10^{21}$~eV. For non-relativistic shocks,
the maximum energy of acceleration $E_p^{\rm max}$ can optimistically
reach $1\times10^{21}$~eV.  For relativistic shocks this
energy can be somewhat higher.  Production of cosmogenic neutrinos
with still higher energies depends on less-developed ideas, such as
acceleration in strong electromagnetic waves, exotic plasma mechanisms of
acceleration and unipolar induction.

The top-down scenarios, on the other hand, naturally provide neutrinos
with energies higher and much higher than $1\times 10^{20}$~eV
\cite{HSW}.  The mechanism common to many models assumes the
existence of superheavy particles with very large masses up to the GUT
scale $\sim 10^{16}$~GeV.  Such particles can be produced by
Topological Defects (TD) (see \cite{CosmStrBookVilenkin} for a general review).
They then rapidly decay and produce a parton cascade, which is
terminated by production of pions and other hadrons.  Neutrinos are
produced in hadron decays.

The production of unstable superheavy particles --- the constituent
fields of TD --- is a very common feature of the TD \cite{BS98}.
However, the dynamics of TD is highly nonlinear and complicated, the
distance between TDs is model-dependent, and the calculation of UHE
particle fluxes requires special consideration for different types of
TD \cite{BBV}.

{\em  Cosmic strings} can release particles in the process of
self-interaction, and in the final evaporation of tiny loops, but only
a few particles are produced by each such interaction.  Of more
interest are cosmic string {\em cusps}, where the string doubles back
on itself and moves with a huge Lorentz factor \cite{McGibbon}.
Particles emitted by cusps have energies much higher than their rest
masses, because of the boost.  However, the flux from such events is
too low to be observed \cite{OBP1, OBP2}.

{\em Monopole-antimonopole pairs connected by strings}
\cite{Hill,Bh,OBP3} can release superheavy particles when the monopole
and antimonopole finally annihilate.  However, such defects, similar
to superheavy dark matter (see below), would be accumulated inside
galaxies, and in particular in the Milky Way.  The resulting UHECR
flux would be dominated by photons, which can reach us easily from
short distances.  Such photons are not observed 
\cite{gamma-Auger} at the level that would be necessary if top-down 
production were to account for the observed UHECR.

If each monopole is attached to two strings, we have {\em necklaces}.
Necklaces are an attractive source for UHE neutrinos \cite{BV,ABK},
but simple models of necklaces may lead to rapid annihilation of the
monopoles \cite{Olum}.  In other models, however, the monopoles may
survive for much longer, providing a detectable flux of UHE
neutrinos.\footnote{The main point of Ref.~\cite{Olum} is that the
relativistic motion of strings causes monopoles to develop large
velocities along the string.  As a result monopoles frequently run
into one another and annihilate.  A possible way to avoid this is to
consider light strings, which remain overdamped till very late times
and therefore move slowly.  Another possibility is that the strings
have zero modes, which act as a one-dimensional gas on the strings and
slow the monopoles down.  These models need further investigation.}

In a wide class of particle physics models, cosmic strings can be {\it
superconducting}, in which case they respond to external
electromagnetic fields as thin superconducting wires \cite{Witten}.
String superconductivity arises when a condensate of charged particles
(which can be either bosons or fermions) is bound to the string.
These particles have zero mass in the bound state, whereas away from
the string they have some mass $m_X$.  Loops of superconducting string
develop electric currents as they oscillate in cosmic magnetic fields.
Near a cusp, a section of string acquires a large Lorentz boost
$\gamma_c$, and simultaneously the string current is increased by a
factor $\gamma_c$.  If the current grows to a critical value $J_{max}$
charge carriers rapidly scatter off each other and are ejected from
the string.
The decay products of these particles can then be observed as
cosmic rays.  This model will be the subject of the present paper.

Apart from TDs, superheavy particles can naturally be produced by
thermal processes \cite{BKV,KR} and by time-varying gravitational
fields \cite{Kolb-grav,Kuz-grav} shortly after the end of inflation.
These particles can survive until present and produce neutrinos in
their decays. Protected by symmetry (e.g.  discrete gauge symmetry, in
particular R-parity in supersymmetric theories), these particles can
have very long lifetimes exceeding the age of the universe.  The
resulting neutrino flux may exceed the observed flux of UHECR.
However, like any other form of CDM, superheavy particles accumulate
in the Milky Way halo and produce a large flux of UHE photons. The
non-observation of these photons puts an upper limit on the neutrino
flux from intergalactic space.

\subsection{The cascade bound}

The neutrino fluxes are limited from above. The most general upper
bound for UHE neutrinos, valid for both cosmogenic neutrinos and
neutrinos from top-down models, is given by the {\em cascade upper
  limit}, first considered in \cite{BS,book}. The production of
neutrinos in these scenarios is accompanied by production of high
energy photons and electrons. Colliding with low-energy target
photons, a primary photon or electron produces an electromagnetic
cascade due to the reactions $\gamma+\gamma_{\rm target} \to e^++e^-$,
$e+\gamma_{\rm target} \to e'+\gamma'$, etc.  The cascade spectrum is
very close to the EGRET observations in the range 3~MeV - 100~GeV
\cite{EGRET}.  The observed energy density in this range is
$\omega_{\rm EGRET} \approx (2 - 3)\times 10^{-6}$~eV/cm$^3$.  To be
conservative, we will use the lower end of this range.  It
provides the upper limit for the cascade energy density.  The upper
limit on UHE neutrino flux $J_{\nu}(>E)$ (sum of all flavors) is given
by the following chain of inequalities \beq \omega_{\rm
  cas}>\frac{4\pi}{c}\int_E^{\infty}E'J_{\nu}(E')dE'>
\frac{4\pi}{c}E\int_E^{\infty}J_{\nu}(E')dE'\equiv
\frac{4\pi}{c}EJ_{\nu}(>E)\,.
\label{eq:int-limit}
\eeq
Here $c$ is the speed of light, but will generally work in units where
$c = 1$ and $\hbar = 1$.  In terms of the differential neutrino
spectrum,  Eq.\ (\ref{eq:int-limit}) gives $J_{\nu}(E)$ gives
\begin{equation}
E^2 J_{\nu}(E) < \frac{c}{4\pi}\omega_{\rm cas },~~ {\rm with}~
\omega_{\rm cas} <\omega_{\rm EGRET}
\label{cas-rig}
\end{equation}

\begin{figure*}[t]
\begin{center}
\mbox{\includegraphics[width=12cm,height=7cm]{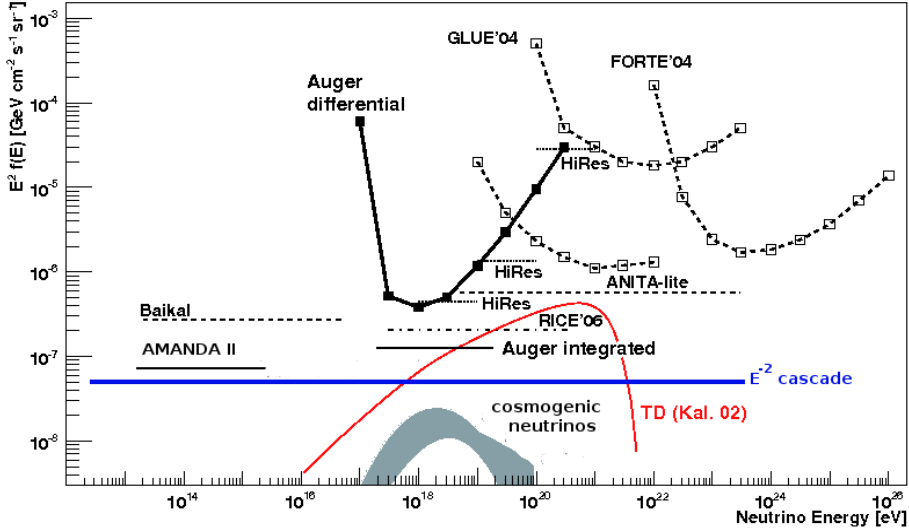}}
\end{center}
\caption{
The experimental upper limits on UHE neutrino fluxes 
in comparison with the electromagnetic cascade upper limit in assumption of 
$E^{-2}$ generation spectrum (labeled ``$E^{-2}$ cascade'') and 
with predictions for cosmogenic neutrinos.
Neutrino fluxes are given for one neutrino flavor $\nu_i+\bar{\nu}_i$.
}
\label{fig:upper-limits}
\end{figure*}

Eq.~(\ref{cas-rig}) gives a {\em rigorous} upper limit on the neutrino flux. 
It is valid for neutrinos produced by HE protons, by topological defects, by
annihilation and decays of superheavy particles, i.~e., in all cases
when neutrinos are produced through decay of pions and kaons.  It holds
for an arbitrary neutrino spectrum decreasing with energy. If one assumes
some specific shape of neutrino spectrum, the cascade limit becomes 
stronger. For a generation spectrum proportional to $E^{-2}$, which is 
used for analysis of observational data, one obtains a stronger upper
limit.  Given for one neutrino flavor it reads \cite{pylos}
\begin{equation}
E^2J_i(E) \leq \frac{1}{3} \frac{c}{4\pi}\frac{\omega_{\rm cas}}
{\ln (E_{\rm max}/E_{\rm min})},
\label{cas-E2}
\end{equation}
where $E_{\rm max}$ and $E_{\rm min}$ give the range of neutrino
energies to which the $E^{-2}$ spectrum extends, and
$i=\nu_{\mu}+\bar{\nu}_{\mu}$, or $i=\nu_e+\bar{\nu}_e$, or 
$i=\nu_{\tau}+\bar{\nu}_{\tau}$. This upper limit is shown in Fig.~\ref{fig:upper-limits}.
One can see that the observations almost reach the cascade upper 
limit and thus almost enter the region of allowed fluxes. 

The most interesting energy range in Fig.~\ref{fig:upper-limits} 
corresponds to 
$E_{\nu} > 10^{21}$~eV, where acceleration cannot provide protons 
with sufficient energy for production of these neutrinos.
At present the region of $E_{\nu} > 10^{21}$~eV, and especially   
$E_{\nu} \gg 10^{21}$~eV is considered as a signature of 
top-down models, which provide these energies quite naturally. 

\subsection{Model assumptions}

In this paper we consider superconducting string loops as a source of
UHE neutrinos.  We consider a simple model in which a magnetic
field of magnitude $B$, occupying a fraction of space $f_B$, is
generated at some epoch $z_{max}\sim$ 2--3.
The strings are characterized by two parameters: the fundamental
symmetry breaking scale $\eta$ and the critical current $J_{max}$.  We
take the mass per unit length of string to be $\mu = \eta^2$.

The predicted flux of UHE neutrinos depends on the typical length
of loops produced by the string network.  This issue has been a
subject of much recent debate, with different simulations
\cite{Shellard,Maria,OVV,OV} and analytic studies \cite{Polchinski,Vitaly} 
yielding different answers.  Here we shall adopt the picture suggested
by the largest and, in our view, the most accurate simulations of
string evolution performed to date
\cite{OVV,OV}.  According to this picture, the characteristic length of
loops formed at cosmic time $t$ is given by the scaling relation
\beq
l\sim \alpha t,
\label{alpha}
\eeq
with $\alpha\sim 0.1$.  

For simplicity and transparency of the formulae obtained in this paper
we use several simplifications. We assume cosmology without $\Lambda$
term with $\Omega_{cdm} + \Omega_b=1$, 
the age of the universe $t_0 = (2/3)H_0^{-1} = 3\times 10^{17}$~s, 
$t_{eq} \sim 1\times 10^{12}$~s,~ and  $(1+z)^{3/2}=t_0/t$
for the connection of age $t$ and redshift $z$ in the matter era.

We also assume the fragmentation function for the decay of
superheavy $X$ particle into hadrons is
\beq\label{E-2}
dN/dE \propto E^{-2},
\eeq
while Monte Carlo simulation and the DGLAP method give closer to
$E^{-1.92}$ \cite{DGLAP}.  

These simplifications give us a great advantage in
understanding the dependence of calculated physical quantities on the
basic parameters of our model, in particular on fundamental string
parameter $\eta$.  
Our aim in this paper is to obtain the order of magnitude of the flux
of UHE neutrinos and to indicate the signatures of the model. We
believe our simplified model assumptions are justified, given the
uncertainties of string evolution and of the evolution of cosmic
magnetic fields.

\section{Particle emission from superconducting strings}

\subsection{Particle bursts from cusps}

As first shown by Witten \cite{Witten}, cosmic strings are
superconducting in many elementary-particle models. As they
oscillate in cosmic magnetic fields, such strings develop electric
currents.
Assuming that the
string loop size is smaller than the coherence length of the field
$l\alt l_{B} \sim 1Mpc$, the electric current can be estimated as
\cite{Witten,CosmStrBookVilenkin}
\begin{equation}
J \sim 0.1 e^2 B l . 
\label{CCurrent}
\end{equation}

Particles are ejected from highly accelerated parts of superconducting
 strings, called cusps, where large electric currents can be induced
\cite{Spergel1,Babul}. The current near a cusp region is boosted as
\begin{equation}
J_{cusp} \sim \gamma_c J,
\end{equation}
where $J$ is the current away from the cusp region and $\gamma_c$ is the
Lorentz factor of the corresponding string segment.  Particles are
ejected from portions of the string that develop Lorentz factors
\begin{equation}
\gamma_c \sim J_{max} \slash J,
\end{equation}
where the current reaches the critical value $J_{max}$.  This maximum
current is model-dependent, but is bounded by $J_{max}\alt e\eta$,
where $\eta$ is the symmetry breaking scale of the string and $e\sim
0.1$ is the elementary electric charge in Gaussian units, renormalized
to take into account self-inductance \cite{CosmStrBookVilenkin}.

One may parametrize $J_{max}$ by introducing the parameter $i_c < 1$:
\beq
J_{max}=i_c e\eta, 
\label{Jmaxeta}
\eeq 

If the charge carrier is a superheavy particle $X$ with mass $m_X$,
the case which will be considered here, one may use $\epsilon_X^r$
for the energy of $X$-particle in the rest system of the cusp and
$\epsilon_X$ in the laboratory system. Then $\epsilon_X^r=\gamma m_X= i_c
\eta$ and
\beq
\epsilon_X \sim i_c \gamma_c \eta, 
\label{E_X}
\eeq
respectively, where $\gamma$ is the average Lorentz factor of 
X-particle in the rest system of the cusp. In Eq.~(\ref{E_X}) we took
into account that the energy of $X$-particle in the laboratory
system is boosted by the Lorentz factor of the cusp $\gamma_c$. 
  
The number of $X$ particles per unit invariant length of the string is
$\sim J\slash e$, and the segment that develops Lorentz factor
$\gamma_c$ includes a fraction $1\slash \gamma_c$ of the total
invariant length $l$ of the loop. Hence, the number of $X$ particles
ejected in one cusp event (burst) is
\begin{equation}
N_{X}^b \sim (J \slash e) (l\slash \gamma_c) \sim J^2 l/eJ_{max}\; .
\label{NX}
\end{equation}
The oscillation period of the loop is $l/2$, so assuming one
cusp per oscillation, the average number of $X$ particles emitted per unit
time is
\beq
\dot{N}_X \sim 2J^2/eJ_{max},
\label{dotN_X}
\eeq
and the luminosity of the loop is
\beq
L_{tot} \sim \dot{N}_X \epsilon_X .
\label{Ltot}
\eeq

The $X$ particles are short-lived. 
They decay producing the parton cascade which is developed due 
to parton splitting in the perturbative regime, until at the
confinement radius the partons are converted into hadrons, mostly pions
and kaons, which then decay producing gamma rays, neutrinos, and electrons.
These particles together with less numerous nucleons 
give the observational signatures of superconducting cusps. 

The neutrino spectrum at present epoch $z=0$, produced by the decay
of one X-particle with energy $\epsilon_X \sim i_c \gamma_c \eta$ at 
epoch $z$ can be calculated using the fragmentation function  
(\ref{E-2}) for an X-particle at rest: 
\beq 
\label{Fragmentation}
\xi_{\nu}(E) \approx \frac{i_{c} \eta \gamma_{c}}{2 (1+z) \ln
  (E_{max}^{rest}/E_{min}^{rest})} \frac{1}{E^{2}} , 
\eeq 
where
$E_{max}^{rest}$ and $E_{min}^{rest}$ are the maximum and minimum
neutrino energies in the rest system of X-particle.

Particle emission from a cusp occurs within a narrow cone
of opening angle 
\begin{equation}
\label{OpeningAngle}
\theta_c \sim \gamma_c^{-1} \sim J \slash J_{max}
\end{equation}  
The duration of a cusp event is \cite{Babul}
\begin{equation}
t_{burst} \sim l \gamma_c^{-3}
\end{equation}

\subsection{Superconducting loops in the universe}

In any horizon-size volume of the universe at arbitrary time there are a
few long strings crossing the volume and a large number of small
closed loops. As loops oscillate under the force of string tension, they
lose energy by emitting gravitational waves at the rate 
\beq
{\dot E}_g\sim \Gamma G\mu^2,
\label{grav-loss}
\eeq
where $\mu \sim \eta^2$ is the string mass per unit length, 
$G=1/m_{Pl}^2$ is the gravitational constant and 
$\Gamma \sim 50$ is a numerical coefficient. 

The number density of loops with lengths in the interval from $l$ to
$l+dl$ at time $t$ can be expressed as $n(l,t)dl$.  Of greatest
interest to us are the
loops that formed during the radiation era $t < t_{eq}$ and still survive at
$t>t_{eq}$.  The density 
of such loops at time $t$ is given by \cite{CosmStrBookVilenkin} 
\begin{equation}
n(l,t)dl \sim t_{eq}^{1\slash 2} t^{-2} l^{-5\slash2}dl,
\label{LoopDensity}
\end{equation}
in the range from the minimum length $l_{min}$ to the maximum 
length $l \sim \alpha t_{eq}$, where 
\beq 
l_{min} \sim \Gamma G \mu t  \sim 3\times 10^{11} \eta_{10}^2 (1+z)^{-3/2}
{\rm cm}
\label{lmin}
\eeq
and $\eta_{10} = \eta/10^{10}$~GeV.  Here and below we assume that the
loop length parameter in (\ref{alpha}) is $\alpha \sim 0.1$, as
suggested by simulations \cite{OVV,OV}.
Loops of the minimum length are of most importance in our calculations
because they are the most numerous.

For a loop of length $l$ at redshift $z$, the Lorentz factor at the cusp
$\gamma_{c}$ can be expressed as
\beq
\gamma_{c} = \frac{J_{cusp}}{J} = \frac{i_{c} e \eta}{0.1 e^2 B l} = 
\gamma_{c} (l_{min}) \frac{l_{min}}{l}
\eeq
where $\gamma_{c} (l_{min}) = \gamma_{0} (1+z)^{3/2}$ and 
\beq
\gamma_{0} = \frac{10 i_{c} \eta}{e B t_{0} \Gamma G \mu}
= \frac{1.1\times 10^{12} i_c}{B_{-6}\eta_{10}} 
\eeq
where $B_{-6}$ is the magnetic field in microgauss.

\subsection{Limits on $\eta$}

The string motion is overdamped at early cosmic times, as a result of
friction due to particle scattering on moving strings.  The
friction-dominated epoch ends at
\beq
t_*\sim (G\mu)^{-2}t_p,
\eeq
where $t_p$ is the Planck time.  In the above analysis we have assumed
that loops of interest to us are formed at $t>t_*$.  The corresponding
condition, 
\beq
\Gamma G\mu t_0/\alpha \agt t_* ,
\eeq
yields
\beq
\eta\agt 10^9 ~{\rm GeV}.
\label{eta*}
\eeq
For strings with $\eta<10^9$~GeV, loops of the size given by
(\ref{lmin}) never form.  Instead, the smallest loops are those that
form at time $t_*$ with length
\beq\label{lminfriction}
l_{min}\sim \alpha t_*\,,
\eeq
and then survive until the present day.

We should also verify that energy losses due to particle emission and to
electromagnetic radiation in recent epochs (after magnetic fields have
been generated) are sufficiently small, so the lifetimes of the loops
(which we estimated assuming that gravitational radiation is the
dominant energy loss mechanism) are not significantly modified.

The average rate of energy loss due to particle emission is 
\beq
{\dot E}_{part}\sim f_B \dot{N}_X \epsilon_X \sim 2 f_B JJ_{max}/e^2\,
\label{Epart}
\eeq
where we have
used Eqs.~(\ref{dotN_X}) and (\ref{E_X}).  The electromagnetic radiation
power is smaller by a factor $e^2\sim 10^{-2}$.  

The factor $f_B$ in Eq.~(\ref{Epart}) is the filling factor -- the
fraction of space filled with the magnetic field.  It gives the
fraction of time that cosmic string loops spend in magnetized regions.
We assume that loop velocities are sufficiently high that they do not
get captured in magnetized cosmic structures (such as galaxy clusters
or LSS filaments).  To justify this assumption, we note that particle
emission can start only after the cosmic magnetic fields are
generated, that is, at $z\sim 3$ or so.  Before that, gravitational
radiation is the dominant energy loss mechanism, and the loops are
accelerated to high speeds by the gravitational rocket effect
\cite{Vachaspati,Hogan}.  The smallest loops of length (\ref{lmin})
have velocities $v\sim 0.1$, certainly large enough to avoid capture.

The particle emission energy rate (\ref{Epart}) should be compared to
the gravitational radiation rate (\ref{grav-loss}).


The ratio of the two rates is zero at $z > z_{max}$, where $z_{max} \sim$
  2--3 is the red-shift of magnetic field production. At $z < z_{max}$ 
it is given by 
\beq
{\dot E}_{part}/{\dot E}_g \sim 50 f_{-3}B_{-6} i_c \eta_{10}^{-1}
\left(\frac{l}{l_{min}}\right)(1+z)^{-3/2}. 
\label{EpEg}
\eeq
where $f_{-3} = f_B/10^{-3}$ and $l_{min}$ is given by (\ref{lmin}).

If particle emission is the dominant energy loss mechanism, then the
lifetime of a loop is
\beq
\tau_{part}\sim \frac{\mu l}{{\dot E}_{part}} \sim \frac{5\eta}{ei_c f_B B}
\sim 0.025 \frac{t_0\eta_{10}}{f_{-3}B_{-6}i_c} . 
\label{taupart}
\eeq
Note that $\tau$ is independent of $l$.  This means that all loops
surviving from the radiation era decay at about the same time.  

For the time being, we shall assume that particle radiation is
subdominant.  We shall discuss the opposite regime in Section II.G.

\subsection{Rate of cusp events}

The rate of observable cusp bursts (i.e., the bursts
whose spot hits the Earth) is given by
\beq
d\dot{N_{b}} = f_{B} \frac{d\Omega} {4\pi} \nu(l, z) dl \frac{dV(z)}{1+z}
\label{BurstRate}
\eeq
where, as before, $f_{B}$ is the fraction of space with magnetic field
$B$, $d\Omega = 2\pi \theta d\theta$ is the solid angle element,
with $\theta$ limited by the angle of cusp emission $\theta_c \sim
1/\gamma_{c}$; $\nu(l, z) = n(l, z)/(l/2)$ is the frequency of the
bursts with $n(l,z)$ given by Eq.~(\ref{LoopDensity}), and 
$dV(z)$ is a proper volume of space limited by redshifts $z$ and $z+dz$,
\beq \label{ProperVolume}
dV(z) = 54 \pi t_{0}^{3} [(1+z)^{1/2}-1]^{2} (1+z)^{-11/2} dz .
\eeq
Integrating Eq.~(\ref{BurstRate}) over $\theta$, $l$ and $z$, 
we obtain
\beq 
\dot{N_{b}} = 54 \pi (t_{eq} t_{0})^{1/2} (\Gamma G \mu)^{-1/2} 
(e/10i_c \eta)^2 \int_{0}^{z_{max}} dz \frac{[(1+z)^{1/2} -
1]^{2}}{(1+z)^{11/4}} f_B(z) B^2(z)  ,
\label{NBf}
\eeq
where $z_{max}$ is the redshift at which the magnetic fields are
generated.  Since the earth is opaque to neutrinos with
the energies we are considering,
only half of these bursts can actually be detected by any given
detector at the surface of the earth or using the atmosphere.

The value of the integral in (\ref{NBf}) depends on one's assumptions
about the evolution of the magnetic field $B$ and of the volume
fraction $f_B$.  This evolution is not well understood.  
If we take these values out of the integral in Eq.~(\ref{NBf})
as the average and characterize them by the effective values of parameters 
$B_{-6}$ and $f_{-3}$ in the range $0<z<z_{max}$, 
then Eq.~(\ref{NBf}) reduces to
\beq \label{NBurst}
\dot{N_{b}} = 2.7 \times 10^{2} \frac{B_{-6}^2 f_{-3}}{i_{c}^2 \eta_{10}^{3}} 
\frac{I}{0.066}\,\,yr^{-1} ,
\eeq
where the integral
\beq
I = \int_{0}^{z^{\prime}} dz \frac{[(1+z)^{1/2} - 1]^{2}}{(1+z)^{11/4}}
= \frac{4}{3} [1 - (1+z^{\prime})^{-3/4}] - \frac{8}{5} [1 - (1+z^{\prime})^{-5/4}] + 
\frac{4}{7}[1 - (1+z^{\prime})^{-7/4}], 
\label{int}
\eeq
is equal to $0.015$, $0.042$ and $0.066$ for 
$z^{\prime} = z_{max} = 1$, $2$ and $3$, respectively. 

The integrand in Eq.~(\ref{NBf}) includes the product $f_B(z)B^2(z)$.
In the calculations of other physical quantities below, similar
integrals will have different combinations of $f_B(z)$ and
$B(z)$. Nevertheless, we shall assume that the average values taken
out of the integral are characterized by approximately the same values
of $f_{-3}$ and $B_{-6}$.

All cosmic structures --- galaxies, clusters, and filaments of the
large-scale structure --- are magnetized and contribute to the rate of
cusp bursts.  In the recent epoch, $z\alt 1$, the dominant
contribution is given by clusters of galaxies with $B_{-6}^{2} f_{-3}
\sim 1$.  The magnetic fields of galaxies have about the same magnitude, 
but the corresponding filling factor $f_B$ is orders of magnitude
smaller.  We shall assume that this holds in the entire interval
$0<z<z_{max}$.  The sources in our model are then essentially clusters
of galaxies.

\subsection{Diffuse flux of UHE neutrinos}

The diffuse differential neutrino flux, summed over all produced
neutrino flavors, is given by the formula
\beq
J_{\nu}(E) = \frac{1}{4\pi} \int d\dot{N_{b}} N_{X}^{b} \xi_{\nu} (E)
\frac{1}{\Omega_{jet} r^2(z)},
\eeq
where $d\dot{N_{b}}$ 
is the rate of cusp bursts (\ref{BurstRate}),
$N_{X}^{b}$ is the number of $X$ particles produced per burst, given
by Eq.~(\ref{NX}),
$\xi_\nu(E)$ is the neutrino spectrum produced by the decay of one
$X$-particle, given by (\ref{Fragmentation}),
\beq
\Omega_{jet} = \pi \theta_c^{2} = \frac{\pi}{\gamma_{c}^{2}},
\eeq
\beq
r(z) = 3 t_{0} [1 - (1+z)^{-1/2}]
\eeq
is the distance between a source at redshift $z$ and the
  observation point at $z = 0$, and 
$\Omega_{jet} r^{2}$ is the area of the burst spot at the Earth from 
a source at redshift $z$.

Using expressions (\ref{LoopDensity}) and (\ref{ProperVolume}), 
and assuming that the product $f_B(z) B(z)$ does not change much in
the interval $0<z<z_{max}$, we obtain\footnote{We note that
numerical simulations of the magnetic field evolution performed by Ryu
et al.~\cite{Ryu} do indicate that the space average of the magnetic
field $\langle B(z)\rangle = f_B(z) B(z)$ remains roughly constant at
$\sim 10^{-9}$~G for $0<z\alt 3$ and decreases at larger values of
$z$. The effective values $B_{-6}$ and $f_{-3}$  could be
different from those in Eq.~(\ref{NBurst}) for the rate of bursts,
but we neglect the possible difference. }
\beq \label{DiffuseFlux}
E^{2} J_{\nu}(E) = \frac{0.3 i_{c} m_{pl} (t_{eq}/t_{0})^{1/2} 
(e B t_{0}^{2})f_B }{7\pi (\Gamma)^{1/2}\, t_{0} (c t_{0})^2\, 
\ln (E_{max}^{rest}/E_{min}^{rest})} [1- (1+z_{max})^{-7/4}].
\eeq
Numerically, this gives for the neutrino flux summed over  neutrino 
flavors
\beq \label{NFlux}
E^{2} J_{\nu}(E) = 6.6 \times 10^{-8} i_{c} B_{-6} f_{-3}\,\,\,\,GeV\,cm^{-2}\,s^{-1}\,sr^{-1},
\eeq
where we have set $z_{max} =3$ and estimated the logarithmic
factor as $\sim 30$.

For $i_c \sim 1$, the flux (\ref{NFlux}) is close to the cascade
upper limit shown in Figure \ref{fig:upper-limits}. Notice that the
diffuse neutrino flux (\ref{DiffuseFlux}) does not depend on
$\eta$. The neutrino flux must correlate with clusters of galaxies.

To detect this flux, we need to monitor a target with some large
mass M.  The effective cross-section of the detector is then
\beq
\Sigma = \sigma_{\nu N} M/m_N
\eeq
where $\sigma_{\nu N} \sim 3\times 10^{-32}\,\,\, cm^{2}$ is the
neutrino-nucleon cross section at $E \agt 10^{10}$~GeV and $m_N$ the
mass of a nucleon.  Because of the opacity of the earth, the
detector will see solid angle about $2\pi$ sr.  The detection rate
of particles with energy above E is
\beq
2 \pi E J_\nu(E) \Sigma \approx 23
\left(\frac{M}{10^{18}g}\right)\left(\frac{10^{10}~GeV}{E}\right) i_{c}
B_{-6} f_{-3}\text{yr}^{-1}
\label{Ndotic}
\eeq
In the
case of JEM-EUSO in tilt mode, $M\sim 5\times10^{18} g$, and thus
we expect about $100 i_c$ detections per year, so events can be
expected for $i_c\agt 0.01$. 

\subsection{Neutrino fluence and the number of neutrinos detected 
from a burst}

The fluence of neutrinos incident on the detector from a burst at
redshift $z$ can be calculated as
\beq
\Phi (>E) =  \frac{N_{X}^{b} \xi_\nu(>E)}{\Omega_{jet} r^2(z)}
\eeq

Consider a neutrino burst from a loop of length $l$ at redshift
$z$. Using $N_{X}^{b}$ from (\ref{NX}), $l_{min}$ from
(\ref{lmin}) and $\xi_{\nu}(>E)$ from (\ref{Fragmentation}), we
obtain for a loop of any length $l$,
\beq 
\Phi (>E) \approx \frac{10 i_{c}^{3} \eta^{3}}{18 \pi e B
  t_{0}^{2}\, E\, \ln (E_{max}^{rest}/E_{min}^{rest}) [(1+z)^{1/2}
    -1]^{2}} ,
\eeq
which numerically results in
\beq \label{NNumber}
\Phi(>E) \approx 1.2\times 10^{-2} \frac{i_{c}^{3}
  \eta_{10}^{3}}{B_{-6}} \left(\frac{10^{10}\,\,GeV}{E}\right)
\frac{1}{[(1+z)^{1/2} -1]^{2}}~ \text{km}^{-2} 
\eeq

The number of neutrinos detected in a burst is 
\beq
N_{\nu}^{det} \sim \Phi(>E)\Sigma
\eeq
With $M\sim 5\times10^{18} g$ as above,
\beq
N_{\nu}^{det}(>E) \approx 0.11 \frac{10^{10}~ {\rm GeV}}{E} 
\frac{i_c^3 \eta_{10}^3} {B_{-6}}
\frac{1}{[(1+z)^{1/2} -1]^2}
\label{Ndet}
\eeq
Therefore, for a certain range of $i_c\eta_{10}$ values and source
redshifts $z$, multiple neutrinos can be detected as parallel tracks  
from a single burst. For example, for $i_c \eta_{10} \sim 3$, and 
$z \sim 1$, ~~ $N_{\nu}^{det} \sim 17$. 

For neutrino energies of interest, $E_{\nu} \agt 1\times 10^{20}$~eV,
the neutrino Lorentz factor is so large that there is practically no
arrival delay for neutrinos with smaller energies. All neutrinos from
a burst arrive simultaneously and produce atmospheric
showers with parallel axes, separated by large distances. 

For other sets of parameters $N_{\nu}^{det} < 1$ , i.e. only one
neutrino from a burst (or no neutrino) is detectable.  As $\eta$ 
increases, the rate of bursts (\ref{NBurst}) diminishes while the
number of neutrinos per burst increases, so that the total neutrino flux
remains unchanged.
\begin{figure*}[t]
\begin{center}
\mbox{\includegraphics[width=11cm,angle=-90]{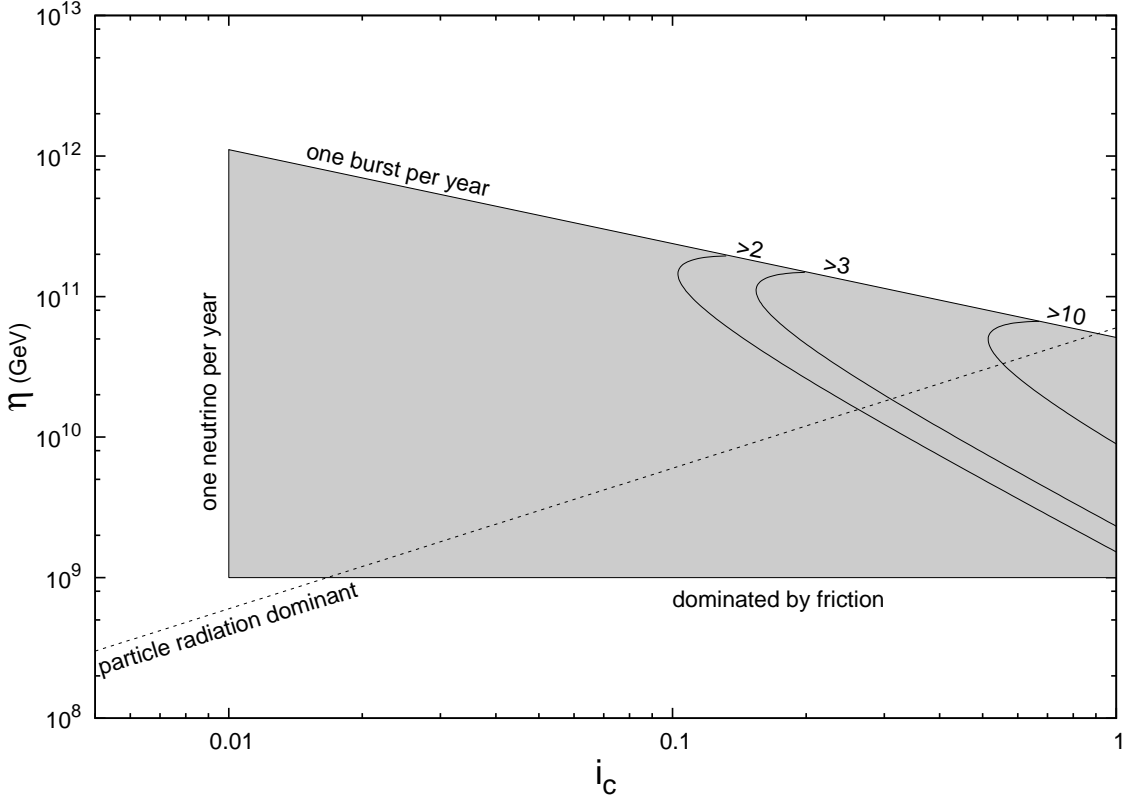}}
\end{center}
\caption{The region of parameter space where neutrinos can be seen by
  a detector with the parameters of JEM-EUSO.  The curved lines show the
  left edges of the regions in which bursts containing at least 2, 3,
  and 10 neutrinos can be expected at least once per year. Below
  the dotted line, particle radiation is the dominant channel of
  energy loss from loops.}
\label{fig:parameters}
\end{figure*}
The rate of detected neutrino bursts with the number of detected
neutrinos $N_{\nu}^{det} > \zeta$ for each burst, is given by 
Eqs~(\ref{NBurst}) and (\ref{int}), with $z_{max}$ determined by 
$N_{\nu}^{det}(>E,z_{max})=\zeta$ .  Using Eq.~(\ref{Ndet}) we obtain 
for $x_{max} \equiv (1+z_{max})$:
\beq
x_{max}(>E,\zeta)= \left [1+ \left (\frac{0.11}{\zeta} 
 \frac{i_c^3\eta_{10}^3}{B_{-6}} \frac{10^{10}~{\rm GeV}}{E} 
\right )^{1/2}\right ]^2, 
\label{Xmax}
\eeq
if (\ref{Xmax}) is less than
4, and $x_{max}=4$ if (\ref{Xmax}) is larger than 4.
Introducing in Eq.~(\ref{NBurst}) coefficient $1/2$ which approximately
takes into account the absorption of UHE neutrinos crossing the
Earth we obtain for the rate of detected bursts with $N_{\nu}^{det}
\geq \zeta$ 
\beq
\dot{N}_b^{det}(\geq \zeta) = 2.1 \times 10^3 \frac{f_{-3} B_{-6}^2}
{i_c^2 \eta_{10}^3} I(z_{max}) ~~ {\rm yr}^{-1}, 
\label{Nbzeta}
\eeq
where $I(z_{max})$ is given by Eq.~(\ref{int}) with $z_{max}$ from 
Eq.~(\ref{Xmax}).  

In Fig.~\ref{fig:parameters}, we have shaded the region of the
parameter space $(\eta,i_c)$ corresponding to a detectable flux of
neutrinos.  Curved lines in the figure mark the regions where we
expect a burst with a given multiplicity of neutrinos, $\zeta =$ 2, 3 or
10, detected simultaneously by a detector with the parameters of
JEM-EUSO tilted.  To the left of the 2-neutrino-burst line, only a
diffuse flux of single neutrinos can be observed.  This flux depends
only on $i_c$, and the vertical left boundary of the shaded region
marks the value of $i_c$ at which it drops below one particle detected
per year.

Note that the regions shown for multiple events are those where
we expect at least one burst per year whose average multiplicity is
the given $\zeta$ or more.  But it is possible even if the parameters
are to the left of the $\zeta = 2$ line that we would happen to
observe multiple neutrinos from a single burst, which would give a
clear signature of neutrino-jet emission from cusps.

Another quantity of interest is the {\em rate of detected neutrinos}
$f_{\nu}(\geq \zeta)$ in the events with neutrino multiplicity greater
than $\zeta$.  It is given by
\beq
f_{\nu}(\geq \zeta)= \frac{1}{2} \int \frac{f_B}{2} \frac{1}{\gamma_c^2}
\frac{n(l,z) dl}{l} \frac{dV(z)}{1+z} N_{\nu}^{det}(>E,z,l).
\label{f_nu1}
\eeq
The important feature of the calculations is the
independence of  $N_{\nu}^{det}(>E,z,l)$ from  $l$. This allows us to
integrate over $l$ in Eq.~(\ref{f_nu1}) to obtain 
\beq
f_{\nu}(\geq \zeta) = 2.1 \times 10^3 \frac{f_{-3} B_{-6}^2}
{i_c^2\eta_{10}^3} \int_0^{z_{max}(\zeta)} dz (1+z)^{-\frac{11}{4}}
\left [ (1+z)^{1/2} - 1 \right ]^2 N_{\nu}^{det} (>E,z) ,
\label{f_nu2}
\eeq
where $z_{max}(\zeta)$ is given by Eq.~(\ref{Xmax}). Using 
Eq.~(\ref{Ndet}) for $N_{\nu}^{det}(>E,z)$ results in 
\beq
f_{\nu}(\geq \zeta) = 1.3 \times 10^2 i_c f_{-3} B_{-6} 
[1 - x_{max}^{-7/4}(i_c,\eta_{10})]~ {\rm yr}^{-1}.
\label{f_nu3}
\eeq
for $E > 1\times 10^{19}$~eV.  The asymptotic expression at  
$0.11 i_c^3 \eta_{10}^3 / B_{-6}\zeta \ll 1$ gives  
\beq
f_{\nu}(\geq \zeta) = \frac{1.5 \times 10^2}{\sqrt{\zeta}} 
i_c^{5/2} \eta_{10}^{3/2} B_{-6}^{1/2}~ {\rm yr}^{-1}
\label{f_nu4}. 
\eeq

\subsection{Neutrino fluxes in the particle-emission dominated regime}

So far we have assumed that gravitational radiation is the
dominant energy loss mechanism of strings.  
In the opposite regime, where the particle emission energy
losses dominate, the loop's lifetime $\tau_{part}$ is 
independent of its length and is given by Eq.~(\ref{taupart}).
We shall analyze this regime in the present section.

As before, we shall adopt the idealized model where the magnetic field
$B$ is turned on at time $t=t_B$, corresponding to redshift $z_{max}$,
\beq
t_B\sim t_0(1+z_{max})^{-3/2} .
\eeq
The loops decay at the time $t_{dec}\sim t_B + \tau_{part}$.  The rate
of observable bursts ${\dot N}_b$ is given by Eq.~(\ref{NBurst}) with
$I$ from Eq.~(\ref{int}), where the integration is taken between
$z_{dec}$ and $z_{max}$ and $z_{dec}$ is the redshift corresponding to
the time $t_{dec}$.

If $\tau_{part} \agt t_B$, the redshift $z_{dec}$ is significantly
different from $z_{max}$, with $\Delta z = z_{max} - z_{dec} \agt
1$, and the value of $I$ is not much different from that evaluated in
Sec. II.D.  This is an intermediate regime, in which the results we
obtained in Sections II.D and II.E for the rate of bursts and for the
diffuse flux can still be used as order of magnitude estimates.

For $\tau_{part}\ll t_B$, the loops lose all their energy to particle
emission in less than a Hubble time.  The condition $\tau_{part}\sim
t_B$ can also be expressed as ${\dot E}_{part}/{\dot E}_g (z_{max})\sim
1$.   Using Eq.~(\ref{EpEg}) with $z_{max}\sim 3$, we find this
condition is met for the smallest loops when
\beq
\eta \sim 6\times 10^{10} i_c f_{-3} B_{-6}~~{\rm GeV}. 
\label{partradcond}
\eeq
It marks the boundary of the strong particle-emission domination
regime and is shown by the inclined dotted line in Fig.~2.  Below this
line, the results of the preceding sections do not apply even by order
of magnitude, but as we shall see, detectable neutrino fluxes can
still be produced.

The redshift interval $\Delta z = z_{max}-z_{dec}$ for
$\tau_{part}\ll t_B$ can be estimated as
\beq
\Delta z \approx \frac{2}{3}\frac{\tau_{part}}{t_B}(1+z_{max}) \ll 1 , 
\label{Deltaz}
\eeq
and the integral $I$ in Eq.~(\ref{int}) is given by 
\beq
I \approx \Delta z {[(1+z_{max})^{1/2}-1]^2 \over{(1+z_{max})^{11/4}}} .
\eeq
With $z_{max}\sim 3$, we have $t_B \sim t_0/8$, and
\beq
{\tau_{part}\over{t_B}}\sim 0.2 {\eta_{10}\over{f_{-3}B_{-6}i_c}} .
\eeq

The rate of bursts that are actually detected, ${\dot N}_b^{det}$,
can be expressed as a product of ${\dot N}_b$ and the probability
$p_\nu^{det}$ that at least one neutrino from the burst will be
detected.  This probability is simply related to the average number of
detected neutrinos per burst $N_\nu^{det}$, given by Eq.~(\ref{Ndet}),
\beq
p_\nu^{det} = 1-\exp(-N_\nu^{det}) .
\eeq
For $N_{\nu}^{det}\ll 1$, we have 
\beq
p_\nu^{det} \approx N_\nu^{det} 
\label{pdet}
\eeq
and again taking $E > 1\times 10^{19}$~eV,
\beq
{\dot N}_b^{det} \sim {\dot N}_b N_\nu^{det} \sim {60
\eta_{10}\over{(1+z_{max})^{7/4}}}~ yr^{-1}
\sim 5\eta_{10} ~yr^{-1},
\label{Ndoteta}
\eeq 
where in the last step we have used $z_{max}\sim 3$.  Requiring that
${\dot N}_b^{det} \agt 1~yr^{-1}$, we obtain the condition
\beq
\eta\agt 10^9 ~GeV.
\label{cond2}
\eeq 
Note that at the boundary of detectability, where $\eta\sim 10^9$ GeV,
we always have $N_\nu^{det}\ll 1$, and thus the approximation
(\ref{pdet}) is justified.  This boundary is the lower horizontal line
bounding the observable parameter range in Fig.~2.  Note also that
Eq.~(\ref{cond2}) coincides with with the condition (\ref{eta*}) for
the burst-producing loops to be unaffected by friction.  

It is interesting to note that the detection rate (\ref{Ndoteta}) in
the particle-emission dominated regime is independent of $i_c$ and
depends only on the symmetry breaking scale $\eta$.  This is in
contrast with Eq.~(\ref{Ndotic}) for the case of gravitational
radiation dominance, where the rate is proportional to $i_c$ and
independent of $\eta$.

\subsection{Cascade upper limit on neutrino flux in the superconducting 
string model}

In Subsection I.C of the Introduction, we gave a very general upper
limit for UHE neutrino flux.  The presence of such a limit does
not contradict the existence of stronger upper limits in some
particular models with additional assumptions. 

In this section, we calculate the energy density of the cascade
radiation in our model and compare it with $\omega_{cas} = 2 \times
10^{-6}\,\,\, eV\,cm^{-3}$ allowed by EGRET measurements.

The cascade energy density can be calculated as
\beq
\omega_{cas} = \int_0^{z_{max}} \frac{dz}{(1+z)^{4}} \int_{l_{min(z)}}^{l_{max(z)}} 
dl f_{B} n(l,t) L_{em}(l,t)
\eeq
where $L_{em}(l, t) \sim \frac{1}{2} L_{tot}(l, t)$ is the loop
luminosity in the form of UHE electrons and photons produced by pion
decays. The standard calculation (for $z_{max}=3$) results in 
\beq \label{CascadeEnergyDensity}
\omega_{cas} \approx \frac{1.2 i_{c} (e B t_{0}^{2}) (t_{eq} / t_{0})^{1/2} 
f_{B} \eta}{7 (\Gamma G \mu)^{1/2} t_{0}^{3}} 
\left [ 1- (1+z_{max})^{-7/4} \right ]
\approx 8.3 \times 10^{-7}
i_c f_{-3} B_{-6}\,\,\, eV\,cm^{-3}
\eeq
The energy density (\ref{CascadeEnergyDensity}) does not depend on
$\eta$ and since $\omega_{cas} < \omega_{EGRET}$, it respects the
general upper limit (\ref{cas-E2}). For $i_c\sim 1$, the
predicted neutrino flux (\ref{NFlux}) is close to the upper limit
shown in figure \ref{fig:upper-limits}.

\section{Gamma-ray jets and single gamma-rays from the cusps}

\subsection{Bursts from loops in the Milky Way}

In each galaxy, including the Milky Way, there are approximately
$N_{l}$ loops with $l \agt l_{min}$,
\beq
N_{l} \sim n(>l_{min}) V_g \sim 2.5 \times 10^{5} \eta_{10}^{-3} 
V_g/10^3 kpc^3 ,
\eeq
where $V_g$ is the volume of the magnetized part of the galaxy.  A narrow
jet of particles emanating from a cusp on such a loop can in
principle hit the Earth. The probability of such a catastrophic event
is very small because of the smallness of solid angle $\Omega_{jet}$
of jet emission.  The number of jets hitting
an area $S$ on the Earth per unit time does not depend on $S$ if
$S\ll\Omega_{jet} r^{2}$, where $r$ is the distance from the
source. This rate is given by

\beq
\dot{N_{b}} = P V_g \int dl \frac{2n(l)}{l} ,
\eeq
where again we assume one cusp per oscillation,  and 
$P = \Omega_{jet}/4\pi = 1/(4 \gamma_{c})^{2}$ is the probability to hit 
the detector. After the
standard calculations, we obtain for 
$V_g \sim 1\times 10^3~ kpc^{3}$,

\beq
\dot{N_{b}} = 1\times 10^{-13} B_{-6}^{2} \eta_{10}^{-3}
i_{c}^{-2}\,\,\, yr^{-1} .
\eeq 

Thus, for particles propagating rectilinearly, the jets from 
cusps in our galaxy are unobservable.

The most important components of the galactic jets are photons and
neutrinos. A photon jet at the highest energies undergoes widening of
the jet angle due to photon absorption in the galactic magnetic field
\cite{Berez70}. Absorption of photons, $\gamma + B \to B + e^{+} +
e^{-}$, is followed by energy loss by electrons and positrons in the
magnetic field, with the emission of synchrotron photons in directions
different from that of the primary photon. This results in the widening
of the solid angle $\Omega_{jet}$ \cite{Berez70}.

The widening of photon jets in the Milky Way is negligible.  This can
be illustrated by a numerical example. The highest energy of a photon
in a jet is $E_{\gamma}^{max} \sim \gamma_{c} \eta \sim 10^{31}
i_{c}/B_{-6}\,\,eV$.  Photons with $E_{\gamma} \geqslant
10^{25}\,\,eV$ are absorbed in galactic magnetic fields. The produced
electrons and positrons with $E_{e} \sim 10^{25} \,\, eV$ have
lifetime $\tau \sim 10^{3}\,\,s$ for synchrotron energy losses and
attenuation length $l_{att} \sim 3\times 10^{13}\,\,cm$. Since the
Larmor radius of such electrons is $r_{L} \sim 3 \times 10^{28}\,\,
cm$, the deflection angle $\theta \sim l_{att}/r_{L} \sim 10^{-15}$ is
of no consequence.

\subsection{Cascade gamma-radiation from Virgo cluster}

As was discussed above, the photon jets from the galactic cusps are
not widening and thus are invisible. For cusps at large distances,
the widening of the photon jet efficiently occurs in the cascading on
CMB photons, $\gamma + \gamma_{CMB} \to e^{+} + e^{-}$, $e +
\gamma_{CMB} \to e^{\prime} + \gamma^{\prime}$ etc., and the source can
be seen in gamma-radiation. As in the case of diffuse cascade
radiation (see section I.C), all primary photons with energy higher
than the absorption energy $\epsilon_{a}$ are absorbed on CMB
radiation and converted into low-energy cascade photons. Thus, cusps
can be seen in $100\,\,GeV - 100\,\,TeV$ gamma radiation, similar to
the sources of UHE protons which can be seen in $TeV$ gamma-radiation
\cite{Blasi}.

The nearest source from which this radiation can be expected is the
Virgo cluster. It is located at distance $r = 18\,\,Mpc$, and the
number of loops within the core of radius $R_{c} \sim 3\,\,Mpc$, where
a magnetic field $B\sim 10^{-6}\,\,G$ can be reliably assumed, can reach
$n_{l} R_{c}^{3} \sim 7 \times 10^{12} \eta_{10}^{-3}$, with the
luminosity in $\gamma\, e^{+}\,e^{-}$ component for each loop
$L_{loop}^{\gamma} \sim 2 \times 10^{29}\,\, i_{c}
\eta_{10}^{3}B_{-6}\,\, erg\,s^{-1}$. Half of this energy goes into
cascade radiation, $L_{cas} \sim 0.5 L_{loop}^{\gamma}$.

The spectrum of the cascade photons at distance $r \sim 20\,\,Mpc$
has two characteristic energies \cite{book}: the absorption
energy $\epsilon_{a} \sim 100\,\,TeV$ and the energy
$\epsilon_{x}$. The latter is the energy of a photon produced by an
electron born in $\gamma + \gamma_{CMB} \to e^{+} + e^{-}$ scattering
by a photon with $E_{\gamma} = \epsilon_{a}$. The energy of this
electron is $E_{e}
\sim 0.5 \epsilon_{a} \sim 50\,\,TeV$ and the second characteristic
energy is $\epsilon_{x} \sim 20\,\,TeV$ for $r
\sim 20\,\, Mpc$.

The spectrum of cascade photons at observation is calculated in \cite{book} as

\beq \label{SpectrumCascade}
J_{\gamma}(E_{\gamma}) = \begin{cases}K(E_{\gamma}/ \epsilon_{x})^{-3/2}, & E_{\gamma} \leqslant \epsilon_{x} \\
K(E_{\gamma}/ \epsilon_{x})^{-2.0}, & \epsilon_{x} \leqslant E_{\gamma} \leqslant \epsilon_{a}\end{cases}
\eeq

The spectrum constant $K$ in (\ref{SpectrumCascade}) can be expressed
in terms of the cascade luminosity $L_{cas}$ and the distance to the
source $r$ as

\beq
K = \frac{L_{cas}}{\Omega_{eff} r^{2}} \frac{1}{\epsilon_{x}^{2} (2 +
\ln (\epsilon_{a} / \epsilon_{x}))}\,,
\eeq
where $\Omega_{eff}$ is the effective solid angle produced by
scattering of cascade electron in extragalactic magnetic field. In
case of full isotropization $\Omega_{eff} \sim 4\pi$. Cascade
luminosity can be estimated as $1/4$ of the total luminosity of cusps
in a cluster, $L_{cas} \sim \frac{1}{4} L_{loop} N_{loop}$. Using
$L_{loop} = 4.4 \times 10^{29} i_{c} \eta_{10}^{3} B_{-6} \,\,\, erg\,
s^{-1}$ and $N_{loop} \sim 2.5 \times 10^{11} \eta_{10}^{-3}$, valid
for a cluster core with $R_{c}\sim 3\,\,Mpc$, one obtains for the flux

\beq \label{FluxVirgo}
J_{\gamma}(>\epsilon_{x}) = \int_{\epsilon_{x}}^{\epsilon_{a}} dE_{\gamma} J_{\gamma}(E_{\gamma}) \sim 1\times 10^{-13} i_{c} B_{-6} (R_{c}/ 3\,\,Mpc)^{3}\,\,\, cm^{-2}\,s^{-1}
\eeq
which is marginally detectable by present telescopes. Note that
$L_{cas}$ and the flux $J_{\gamma}$ do not depend on $\eta$. We
consider the estimate (\ref{FluxVirgo}) as a very rough indication of
detectability of the gamma-ray flux from the Virgo cluster. Much
more accurate calculations are needed for a reliable prediction of
this flux.

\section{UHE protons from superconducting strings}

The cusps of superconducting strings in clusters of galaxies produce
UHE nucleons at fragmentation of parton jets with a fraction of
nucleons $\epsilon_{N} = 0.12$ \cite{ABK} relative to the total number
of hadrons.  The generation rate $Q_{p}(\Gamma_{p})$ of UHE protons
with Lorentz factor $\Gamma_{p}$ per unit comoving volume and unit
time can be expressed through emissivity,

\beq
\mathcal{L}_{0} = \int_{\Gamma_{p}^{min}}^{\Gamma_{p}^{max}} 
d\Gamma_{p} m_{N} \Gamma_{p} Q_{p}(\Gamma_{p})\,,
\eeq
where the emissivity $\mathcal{L}_{0}$ is the energy released in UHE
protons at $z=0$ per unit comoving volume per unit time,
$\Gamma_{p}^{max}$ and $\Gamma_{p}^{min} \sim 1$ are the maximum and
minimum Lorentz factors of the protons, respectively, and $m_{N}$ is the
nucleon mass. For a power-law generation spectrum $Q_{p}(\Gamma_{p})
\sim \Gamma_{p}^{-2}$, we have

\beq \label{GenerationRate}
Q_{p}(\Gamma_{p}) = \frac{\mathcal{L}_{0}}{m_{N}\, \ln
\Gamma_{p}^{max}} \Gamma_{p}^{-2} .
\eeq

The emissivity is calculated as

\beq
\mathcal{L}_{0} = \epsilon_{N} f_{B} \int_{l_{min}}^{l_{max}} dl n(l) 
L_{tot}^{cusp}(l) ,
\eeq
where $l_{min}$ is given by (\ref{lmin}), while $n(l)$ and $L_{cusp}$
are given by (\ref{LoopDensity}) and (\ref{Ltot}) respectively. For
$L_{tot}^{cusp}$ one readily obtains

\beq
L_{tot}^{cusp} = \frac{J^{2} l}{e J_{c}} \frac{i_{c} \gamma_{c} \eta}{l/2} =
0.2 i_{c} e B l \eta ,
\eeq
and after a simple calculation we have
\beq
\mathcal{L}_{0} \approx 0.4 i_{c} \epsilon_{N} f_{B}
\frac{(t_{eq}/t_{0})^{1/2} e B t_0^2}{(\Gamma G \mu)^{1/2}}  
\frac{\eta}{t_0} \frac{1}{(t_0)^3}  
 \approx 1.4 \times 10^{45} i_{c} f_{-3} B_{-6}\,\,\,erg\,Mpc^{-3}\,yr^{-1} .
\eeq

One more parameter relevant for the calculation of $Q_{p}(\Gamma_{p})$ is
$\Gamma_{p}^{\rm max} = E_{p}^{max} / m_{N}$. It can be estimated using
$E_{p}^{max} \sim 0.1 \epsilon_{X}$, where 
$\epsilon_{X}=i_c\gamma_c \eta$ is the 
energy of the boosted $X$ particles in the laboratory system, 
which being estimated for loops of length $l_{min}$, gives
\beq
\Gamma_{p}^{max} = 1 \times 10^{10} \eta_{10} i_{c}^{2}
\frac{1}{\Gamma G \mu}\frac{\eta}{eBt_0}  \left(\frac{1\,\,GeV}{m_{N}}\right) .
\eeq

Notice that $\Gamma_{p}^{max}$ does not depend on $\eta$ and that it 
enters $Q_{p}(\Gamma_{p})$ through $\ln \Gamma_{p}^{max}$.

Now we can calculate the space density of UHE protons using the
generation 
rate $Q_{p}(\Gamma_{p})$ given by (\ref{GenerationRate}) and taking 
into account propagation through CMB radiation with the help of the
kinetic equation \cite{Longaire,BGGprd}

\beq \label{Kinetic1}
\frac{\partial}{\partial t} n_{p}(\Gamma_{p}, t) - 
\frac{\partial}{\partial_{\Gamma_p}}[b(\Gamma_{p}, t) n_{p}
(\Gamma_{p}, t)] = Q_{p}(\Gamma_{p}, t) ,
\eeq
where $b(\Gamma_{p}, t) = -d\Gamma/dt$ describes energy losses of UHE
protons interacting with CMB photons. For $\Gamma \geqslant 3 \times
10^{10}$, the proton energy losses become large and one can neglect
the first term in the lhs of equation (\ref{Kinetic1}). Then
Eq.~(\ref{Kinetic1}) becomes stationary and its solution for $t =
t_{0}$ reads

\beq
n_{p}(\Gamma_{p}) = \frac{1}{b(\Gamma_{p})}
\int_{\Gamma_{p}}^{\Gamma_{p}^{max}} Q_{p}(\Gamma_{p}) d\Gamma_{p} \approx 
\frac{\mathcal{L}_{0}}{m_{N} \Gamma_{p}\, b(\Gamma_{p})\, 
\ln\Gamma_{p}^{max}} .
\eeq

In terms of the proton energy $E = m_{N}\Gamma_{p}$ and the diffuse
flux $J_{p}(E) = (c/4\pi) n_{p}(E)$, we have, in the standard form of
presentation,

\beq
E^{3} J_{p}(E) \approx \frac{c}{4\pi} \frac{\mathcal{L}_{0}}{\ln \Gamma_p^{max}}
\frac{E^{2}}{b(E)} ,
\eeq
where $b(E)=dE/dt$.
With $b(E)$ taken from \cite{BGGprd} a  numerical  estimate at 
$E = 3\times 10^{19}\,\,\,eV$ gives

\beq \label{PFlux}
 E^{3} J_{p}(E) \approx 1.3 \times 10^{24} i_{c} f_{-3} B_{-6}\,\,\,
 eV^{2}\,m^{-2}\,s^{-1}\,sr^{-1} . 
\eeq
 
With $i_c\sim1$, the calculated flux (\ref{PFlux}) coincides well with
the measurements at the same energy, e.g., with the HiRes \cite{HiRes} 
flux $ E^{3}J_{p}(E) = 2.0 \times 10^{24}\,\,\, 
eV^{2}\,m^{-2}\,s^{-1}\,sr^{-1}$,
so the cusp emission may account for the observed events at the
highest energies. For $i_c \alt 0.1$ the UHE proton flux from
superconducting strings is subdominant. 
 
The UHE proton spectrum from superconducting strings has a sharper GZK
cutoff than the standard spectrum for homogeneously distributed
sources.  This is due to the absence of clusters of galaxies in the
vicinity of our galaxy. The nearest cluster, Virgo, is located at
$18\,\,Mpc$ from the Milky Way; other clusters are located at much
larger distances. Nearby sources affect the spectrum at $E \geqslant 1
\times 10^{20}\,\,eV$, where the proton spectrum from superconducting
strings is predicted to be steeper than the standard one. The
experimental data at present have too low statistics to distinguish
the two cases.
 
In contrast, homogeneously distributed sources such as necklaces
\cite{BV}, give the dominant contribution at $E \geqslant (7-8) \times
10^{19}\,\,eV$ in the form of UHE photons, coming from nearby
sources. In the case of superconducting strings such component is
absent. The UHE photon component from superconducting strings is not
dominant at energy lower than $5 \times 10^{19}\,\,eV$, because
absorption of photons at these energies is stronger than for protons.

\section{Conclusions}

Superconducting cosmic strings produce high energy particles in the
decay of charge carriers, $X$ particles, ejected from the string
cusps. The large Lorentz factor $\gamma_{c}$ of the cusp boosts the
energies of these particles and collimates them in a narrow beam with
opening angle $\theta \sim 1/\gamma_{c}$. The basic string parameter
is $\eta$, the scale of symmetry breaking, which we parametrize as
$\eta = \eta_{10} 10^{10}\,\,GeV$. Another free parameter $i_{c}
\alt 1$ determines the critical electric current in the cusp,
$J_{max} = i_{c} e \eta$, and the mean energy of the charge carriers $X$
escaping from the string, $\epsilon_{X} = i_{c} \gamma_c \eta$.

The astrophysical parameter which determines the electric current
induced in the string is the magnitude of the magnetic field $B$
in the relevant cosmic structures.
The fraction $f_{B}$ of the universe occupied by magnetic field $B$
determines the flux of high-energy particles produced by
superconducting strings. The most favorable values of $B$ and $f_{B}$
for the generation of a large flux of UHE neutrinos are $B \sim
10^{-6}\,\,G$ and $f_{B} \sim 10^{-3}$. They correspond to clusters of
galaxies.

The main uncertainties of our model are related to the uncertainties
in our understanding of the evolution of cosmic strings and of the
origin and evolution of cosmic magnetic fields.  On the cosmic string
side, the key unknown quantity is the parameter $\alpha$ which sets
the characteristic length of string loops in Eq.~(\ref{alpha}).  Here,
we used the value of $\alpha\sim 0.1$, as suggested by numerical
simulations in Refs.~\cite{OVV,OV}.  We have also disregarded the
effects of loop fragmentation.  Toward the end of its life, the
loop's configuration may be sufficiently modified by radiation
back-reaction that the loop will self-intersect and break up into
smaller pieces.  These smaller loops will have more frequent cusps,
shorter lifetimes, higher velocities, and smaller induced currents.
The effect of such loops on the neutrino fluxes is hard to assess without a quantitative model of loop fragmentation.  This will have to await the
next generation of string evolution simulations.  

On the astrophysical side,
basically unknown is the cosmological evolution of the magnetic field
parameters $f_B(z)$ and $B(z)$ in the redshift interval $0 < z <
z_{max}$, where $z_{max} \sim$ 2 -- 3 is the redshift when the magnetic
field was generated. For the space average value $\langle f_B(z) B(z)
\rangle$ we use the numerical simulation by Ryu et al. \cite{Ryu},
according to which this value remains roughly constant at $0 < z <
3$. Some important quantities, such as the diffuse neutrino flux
$J_{\nu}(E)$, the cascade energy density $\omega_{cas}$, and the UHE
proton emissivity are determined by the evolution of the product
$f_B(z)B(z)$.  However, some other quantities, such as the rate of
neutrino bursts and fluence depend on the evolution of $f_B(z)$ and
$B(z)$ in other combinations. In these cases we consider the
parameters $f_{-3}$ and $B_{-6}$ as effective values, using $f_{-3}
\sim B_{-6} \sim 1$.  

In addition, we adopted the following simplifying assumptions. The
Lorentz factor of the cusp is characterized by a single fixed value
$\gamma_{c}$, while in reality there is a distribution of Lorentz
factors along the cusp. The spectrum of particles in a jet is 
approximated as $E^{-2}$, while a QCD calculation
\cite{DGLAP} gives a spectrum which is not a power law, with 
the best power-law fit as $E^{-1.92}$.  
We use cosmology with $\Lambda = 0$. The spectrum of photons from
Virgo cluster and the diffuse spectrum of UHE protons are calculated
using very rough approximations.  Given the uncertainties of
string and magnetic field evolution, these simplifications are rather
benign.  On the other hand, they have the advantage of yielding
analytic formulae, which allow us to clearly see the dependence of the
results on the parameters involved in the problem.  In particular,
with the assumed particle spectrum $\sim E^{-2}$, the diffuse
flux of neutrinos, the cascade upper limit, the flux of $TeV$ photons
from Virgo cluster and the diffuse flux of UHE protons do not depend on
$\eta$. Since the realistic spectrum is very close to $E^{-2}$, this
means that the quantities listed above depend on $\eta$ very weakly.

We summarize the results obtained in this work as follows.

As our calculations show, among different sources, such as galaxies,
group of galaxies, filaments, etc., the largest diffuse flux is
produced by clusters of galaxies with $B\sim 10^{-6}\,\,G$ in a
cluster core and $f_{B} \sim 10^{-3}$.  The calculated diffuse
neutrino flux for three neutrino flavors and for $z_{max}=3$ is

\beq \label{NFlux2}
E^{2} J_{\nu} (E) \sim 6.6 \times 10^{-8} i_{c} f_{-3}
B_{-6}\,\,\,GeV\,cm^{-2}\,s^{-1}\,sr^{-1}.
\eeq

This flux respects the cascade upper limit, provided by the energy
density of electrons, positrons and photons, which initiate
electromagnetic cascades in collisions with CMB photons. The cascade
energy density is calculated from Eq.~(\ref{NFlux2}) as
\beq \label{CascadeEnergyDensity2}
\omega_{cas} \approx 8.3 \times 10^{-7}i_c f_{-3} B_{-6}\,\,\, eV\,cm^{-3} .
\eeq
and is close to the cascade limit for $i_c\sim 1$. 
It is the same as given by Eq.~(\ref{CascadeEnergyDensity}).

At energies $E \alt 10^{22}\,\,eV$, the flux (\ref{NFlux2}) is
detectable by future detectors JEM-EUSO and Auger
(South + North).  The signature of the superconducting string model is
the correlation of neutrinos with clusters of galaxies.  We note,
however, that the neutrino flux from the nearest cluster, Virgo, is
undetectable by the above-mentioned detectors.

Another signature of the model is the possibility of multiple
events, when several showers appear simultaneously in the field of
view of the detector, e.g. JEM-EUSO. They are produced by neutrinos from
the same jet.  The time delay in arrival of neutrinos with different
energies is negligibly small. Such multiple events are expected
to appear for a certain range of parameters, as indicated in Fig.~2.

As an illustration, in Table \ref{table1} we show, for a representative 
value $\eta = 5\times 10^{10}$~GeV, the diffuse neutrino flux,
in units of the cascade upper limit $J_{\nu}^{max}$, 
the rate of bursts, and the average shower multiplicity for several
values of $i_c$.  Note that the bottom row in the table is the {\it
average} multiplicity, that is, the average number of neutrinos
detected per burst.  For example, the low multiplicity at $i_c=0.1$
indicates that only a small number (about 5) out of the 220 bursts per
year will actually be detected.  For $i_c=1/3$, the average
multiplicity is below 1, but Fig.~2 shows that we can expect at least
one 2-neutrino burst per year.

\begin{table}[ht]
\begin{center}
\caption{The diffuse flux $J_{\nu}(E)$ in units of the cascade upper limit 
$J_{\nu}^{max}$ for 3 neutrino flavors, found from (\ref{cas-E2}), 
the rate of neutrino bursts, and the shower multiplicity
(the average number of neutrinos detected in one bursts), for 
$\eta=5\times 10^{10}~$GeV, $z_{max}=3$ and different values of $i_c$.  The
multiplicity is shown for neutrinos with $E \agt 10^{10}$~GeV from a
burst at $z=2$.
}
\vspace{3mm}
\begin{tabular}{c|c|c|c|c}
\hline 
$i_c$   & $1.0$ & $1/2$ & $1/3$ & $0.1$  \\
\hline
$J_{\nu}/J_{\nu}^{max}$ & $0.42$ & $0.21$ & $0.14$ & $0.042$ \\
\hline
rate of bursts & $2.2~{\rm yr}^{-1}$ & $8.7~{\rm yr}^{-1}  $ &
$19.6~{\rm yr}^{-1} $ & $220~{\rm yr}^{-1} $ \\
\hline
multiplicity  & $26$ & $3.2$ & $0.95$ & 0.026 \\
\hline
\end{tabular}
\label{table1}
\end{center}
\end{table}

A photon jet from the cusp initially propagates together with
the neutrino jet, within the same solid angle. However, at large
enough distance, photons from the jet can be absorbed in collisions
with CMB photon ($\gamma + \gamma_{CMB} \to e^{+} + e^{-}$), the
produced electrons (positrons) emit high-energy photons in
inverse-Compton scattering ($e + \gamma_{CMB} \to e^{\prime} +
\gamma^{\prime}$), and thus an em cascade
develops. Electrons are deflected in magnetic fields, and photon
radiation is isotropized. Due to this effect, $10-100\,\,TeV$ gamma
radiation from the nearby cluster of galaxies, Virgo, can be marginally
detectable. The corresponding photon flux 
is given by

\beq \label{FluxVirgo2}
J_{\gamma}(>\epsilon_{x}) \sim 1\times 10^{-13} i_{c} B_{-6}\,\,\,
cm^{-2}\,s^{-1}
\eeq
where $\epsilon_{x} \sim 20\,\,TeV$. 

In the Milky Way, there may be a large number of loops with radiating
cusps, but because of the very small jet opening angle, the 
probability to observe UHE particle jets coming from these loops is
extremely small.

The diffuse flux of UHE protons is suppressed by the small fraction of
nucleons produced at decay of $X$ particles (the factor $\epsilon_{N}
= 0.12$ is obtained in MC and DGLAP calculations \cite{DGLAP}), 
and by energy losses of protons interacting with the CMB during
propagation. 
The calculated flux at energy $E \geqslant 3\times 10^{19}\,\,eV$ 
is given by the approximate formula
\beq \label{PFlux2}
E^{3} J_{p}(E) \approx \frac{c}{4\pi}\frac{\mathcal{L}_{0}}
{\ln \Gamma_{p}^{max}} \frac{E^{2}}{b(E)}
\eeq
where $b(E) = - dE/dt$ is the energy loss rate of protons,
$\Gamma_{p}^{max}$ is the maximum Lorentz factor of a proton at
production, and $\mathcal{L}_{0}$ is the emissivity (energy in the form
of protons emitted per unit comoving volume per unit time),
given by

\beq \label{Emissivity2}
\mathcal{L}_{0} \approx 1.4 \times 10^{45} i_{c} f_{-3} B_{-6}\,\,\,erg\, Mpc^{-3}\,yr^{-1}
\eeq

For $i_c\sim 1$ and $E \sim 3\times 10^{19}\,\,eV$, the proton flux
can reach the value $1.3\times
10^{24}\,\,eV^{2}\,m^{-2}\,s^{-1}\,sr^{-1}$, which can be compared for
example with $2 \times 10^{24}\,\,eV^{2}\,m^{-2}\,s^{-1}\,sr^{-1}$
measured by HiRes \cite{HiRes}.  Thus, radiation from cusps may account for
observed events at the highest energies.  The predicted spectrum at
$E> 8\times 10^{19}\,\,eV$ is steeper than the standard UHECR spectrum
with homogeneous distribution of sources. The accompanying UHE gamma
radiation is very low, due to large distances between the sources
(clusters of galaxies).

As already mentioned, practically all predicted quantities, such
as the diffuse neutrino flux (\ref{NFlux2}), the cascade energy
density (\ref{CascadeEnergyDensity2}), the UHE gamma-ray flux from
Virgo cluster (\ref{FluxVirgo2}), the diffuse flux of UHE protons
(\ref{PFlux2}) and  the proton emissivity (\ref{Emissivity2}),
do not depend on the basic string parameter $\eta$. There are
only two observable quantities that do,
the rate of neutrino bursts $\dot{N_b}$ and the
neutrino fluence $\Phi (>E)$:
\beq
\dot{N_{b}} \sim 3\times 10^{2} \frac{B_{-6}^{2} f_{-3}}{i_{c}^{2} \eta_{10}^{3}}
\,\,yr^{-1}
\eeq 
\beq
\Phi(>E) \approx 1\times 10^{-2} \frac{i_{c}^{3}
  \eta_{10}^{3}}{B_{-6}} \left(\frac{10^{10}\,\,GeV}{E}\right)
\frac{1}{[(1+z)^{1/2} -1]^{2}} \text{km}^{-2}, 
\eeq

As $\eta$ decreases (at a fixed $i_c$), the rate of neutrino
bursts goes up and the number of neutrinos detected in a burst,

\beq
N_{\nu}^{det}(>E) \approx 0.11 \frac{10^{10}~ {\rm GeV}}{E} 
\frac{i_c^3 \eta_{10}^3} {B_{-6}}
\frac{1}{[(1+z)^{1/2} -1]^2}
\eeq
goes down, while the product $\dot{N_{b}} N_{\nu}^{det}$ remains
$\eta$-independent.

We have considered here only ``regular'', field theory cosmic strings.
Recent developments in superstring theory suggest
\cite{Tye,Copeland,Dvali} that the role of cosmic strings can also be
played by fundamental (F) strings and by $D$-branes.  Such strings may
be superconducting, in which case they will also emit bursts of
relativistic particles from their cusps.  The main difference from the
case of ordinary strings is that the probability for two intersecting
strings to reconnect, which is $p=1$ for ordinary strings, can be
$p<1$ and even $p\ll 1$ for F or D-strings.  A low reconnection
probability results in an enhanced density of loops; the particle
production by loops is increased correspondingly.

UHE neutrinos from superconducting strings may have three
important signatures: correlation with clusters of galaxies, multiple
neutrino-induced showers observed simultaneously in the field of view
of a detector, e.g. JEM-EUSO, and detection of $\sim 10\,\,TeV$
gamma-radiation from Virgo, the nearest cluster of
galaxies.

\section{Acknowledgments}

We would like to thank J. J. Blanco-Pillado for useful discussions,
and A. Gazizov for preparing Fig.~1 and discussions.  This work was
supported in part by the National Science Foundation under grants
0353314 and 0457456 (USA), and by the contract ASI-INAF I/088/06/0 for
theoretical studies in High Energy Astrophysics (Italy).

\end{document}